\def\addr#1{{\small\it #1}}

\def\tento#1{\times10^{#1}}
\def\aet#1#2{\approx #1 \tento{#2}}

\def\crr{\cr\noalign{\vskip2pt}}

\def\etal{{\frenchspacing\it et al.}}
\def\ie{{\frenchspacing\it i.e.}}
\def\eg{{\frenchspacing\it e.g.}}

\def\beq#1{\begin{equation}\label{#1}}
\def\eeq{\end{equation}}
\def\beqa#1{\begin{eqnarray}\label{#1}}
\def\eeqa{\end{eqnarray}}
\def\eq#1{equation~(\ref{#1})}

\def\eqnum#1{~(\ref{#1})}

\def\bfig{\begin{figure}[h] \centerline{\hbox{}}\vfill}
\def\efig{\end{figure}\vfill\newpage}

\def\pp{\noindent\parshape 2 0truecm 13.6truecm 1truecm 12.6truecm}
\def\rf#1;#2;#3;#4 {\par\pp#1, {\it #2}, {\bf #3}, #4. \par}
\def\rg#1;#2;#3;#4;#5 {\par\pp#1, {\it #2}, {\bf #3}, #4 (``#5"). 
\par}

\def\Ob{\Omega_b}

\def\izi{\int_0^{\infty}}

\def\erf{\hbox{erf}}

\def\Ms{M_{\odot}}     \def\Ls{L_{\odot}}
\def\Rd{\dot R} \def\pd{\dot p}  
\def\Rdd{\ddot R} \def\md{\dot m}
\def\Ob{\Omega_b}   \def\Od{\Omega_d}
\def\rb{\rho_b}     \def\rd{\rho_d}   \def\rc{\rho_c}
\def\ri{\rho_i}     \def\rs{\rho_s} 
\def\sr3{\sqrt 3}

\def\rdr{{\Rd \over R}}
  
\def\fsn{f_{sn}}  
\def\fm{f_m}
\def\fg{f_g}
\def\tb{t_{burn}}
\def\taub{\tau_{burn}}
\def\Lsn{L_{sn}} \def\Lc{L_{comp}} \def\Lh{L_{diss}}
\def\Lb{L_{brems}} \def\Li{L_{ion}}
\def\l{\ell} \def\h{\eta}
\def\ls{\ell_{sn}} \def\lc{\ell_{comp}} \def\lh{\ell_{diss}}
 \def\li{\ell_{ion}}
\def\Et{E_t}  
\def\rydberg{E_0}
\def\ff{\phi}
\def\r{r}
\def\p{q}
\def\P{P}
\def\t{\tau}
\def\et{\varepsilon_t}  \def\ek{\varepsilon_k}  \def\e{\varepsilon}
\def\ein{\varepsilon_{in}}
\def\fcoll{f_d}

\def\Tg{T_{\gamma}}  
 
\def\st{\sigma_t}

\def\ii{\chi}   
  \def\Tigm{T_{IGM}}
\def\lion{\lambda_{ion}} \def\lrec{\lambda_{rec}}
\def\Lion{\Lambda_{ion}} \def\Lrec{\Lambda_{rec}}
\def\omdc {(1-\delta)^3}

\def\hubpar{\left(\r'-{2\over 3}\h\r\right)}
\def\linj{\ell_{inj}}  \def\finj{f_{inj}}
\def\Tigm{T_{IGM}}
\def\ergs{{\rm ergs}}
\def\Mpc{{\rm Mpc}}

\documentstyle[11pt,epsf,rotate]{article}

\begin{document}


\begin{titlepage}   

\noindent

\begin{center}

\vskip0.9truecm
{\bf

LATE REIONIZATION BY SUPERNOVA DRIVEN WINDS\footnote{
Published in {\it ApJ}, {\bf 417}, 54, November 1, 1993.\\
Submitted September 9 1992, accepted March 24.
Available from\\
{\it h t t p://www.sns.ias.edu/$\tilde{~}$max/gp.html} 
(faster from the US) and from\\
{\it h t t p://www.mpa-garching.mpg.de/$\tilde{~}$max/gp.html} 
(faster from Europe).\\
}
}

\vskip 0.5truecm

Max Tegmark$^1$, 
Joseph Silk$^2$ 
\&
August Evrard$^3$

\smallskip
\addr{$^1$Department of Physics, University of California, 
Berkeley, California  94720}\\
\addr{$^2$Departments of Astronomy and Physics, and
Center for Particle Astrophysics, University of California, 
Berkeley, California 94720}\\
\addr{$^3$Department of Physics, University
of Michigan, Ann Arbor, Michigan 48109}\\

\smallskip
\vskip 0.2truecm

\end{center}

\begin{abstract}
A model is presented in which supernova-driven winds
from early galaxies reionize the intergalactic medium by $z=5$.
This scenario can explain the observed absence of a
Gunn-Peterson trough in the spectra of high-redshift
quasars
providing that the bulk of these early galaxies are quite small,
no more massive than about $10^8 M_{\odot}$. It also
predicts that
most of the IGM was enriched to at least $10\%$ of current
metal content by $z=5$ and perhaps as early as $z=15$.
The existence of such
early mini-galaxies violates no spectral constraints
and is
consistent with a pure CDM model with $b\leq 2$. Since the final
radius of a typical ionized bubble is only around 100 kpc, the
induced modification of the galaxy autocorrelation function is
negligible, as is the induced angular smoothing of the CBR.
Some of the gas swept up by shells may be observable as
pressure-supported Lyman-alpha forest clouds.
\end{abstract}
\end{titlepage}

\section{Introduction}
\label{gpsec1}

The absence of a Gunn-Peterson trough in the spectra of
high-redshift quasars has provided strong
evidence for the intergalactic medium (IGM) being highly ionized as
early as $z=4$ (Gunn \& Peterson 1965; Steidel and Sargent
1987; Webb {\etal} 1992).  The hypothesis that
photoionization of the IGM by quasars could
account for this ionization (Arons \& McCray 1969; Bergeron \&
Salpeter 1970; Sherman 1980) has been challenged
(Shapiro 1986; Shapiro \& Giroux 1987;  Miralda-Escude \& Ostriker
1990). 
Other studies have
maintained that photoionization by quasars (Donahue \& Shull 1987)
or active galactic nuclei (Teresawa 1992) may nonetheless be
sufficient. However, in view of the large uncertainties in crucial
parameters such as ionizing fluxes, the issue of what reionized the
IGM must still be considered open.

In comparing the Gunn-Peterson constraints with our work in papers 3 and 4,
the crucial difference is the degree of ionization required. 
To affect the CBR, it does not really matter whether
the ionization fraction $x$ is $90\%$ or $99.999\%$, as this 
makes only a $10\%$ difference in the optical depth $\tau$.
The Gunn-Peterson limits constrain not $x$ but
$(1-x)$, the neutral fraction. Thus in this context, the difference between
$90\%$ and $99.999\%$ ionization is four orders of magnitude.
In papers 3 and 4, we found that photoionization
by early galaxies could easily reionize the IGM by a redshift 
$z=5$, but the issue here is whether photoionization alone can
provide the extremely high ionization fraction required to pass the Gunn-
Peterson test. 

In this paper, we investigate an alternative  
reionization scenario, which produces considerably higher IGM
temperatures than those attained by the photoionization models in 
previous papers. 
Supernova driven winds from luminous galaxies have long been conjectured
to be an important ionization source for the IGM (Schwartz
{\etal} 1975; Ikeuchi \& Ostriker 1986; Carlberg \& Couchman 1989).
Cold dark matter (CDM)-based models of structure formation
(Blumenthal {\it et al.} 1984; Efstathiou {\it et al.} 1985)
predict the formation of gravitationally bound objects of mass as
small as $10^7\,\Ms$ in large numbers before $z=5$. Recent work
(Blanchard {\it et al.} 1992) indicates that such objects can cool
rapidly and presumably fragment into stars.  These early mini-galaxies
would be expected to release great amounts of kinetic energy into the
surrounding IGM, thereby creating
large, fairly spherical voids filled with thin, hot, ionized
plasma.
We analyze the effect of expanding bubbles
driven by supernova winds from early mini-galaxies, and show that this
mechanism of distributing energy can indeed provide the required
ionization without violating any of the current spectral constraints.

\noindent
In 
Section~\ref{gpsec2},
we will treat the expansion of a shell in a
uniform, cold and neutral IGM. As these bubbles become
larger and more numerous and fill most of space, this obviously
becomes a very poor model of the IGM. In 
Section~\ref{gpsec3}
we estimate bulk
properties of this new processed IGM such as temperature, density
and ionization.

\section{The Explosion Model}
\label{gpsec2}

Since the pioneering work on spherically symmetric explosions by
Sedov (1959), a profusion of analytic solutions have been given by
numerous authors for models of ever-increasing complexity (Cox
\& Smith 1974; McKee \&
Ostriker 1977; Weaver {\it et al.} 1977; McCray \& Snow 1979; Bruhweiler
{\it et al.} 1980; Tomisaka {\it et al.} 1980; McCray \&
Kafatos 1987; Ostriker \& McKee 1988). Most of these models pertain
to bubbles in the interstellar medium of a galaxy, where the expansion
of the universe can be ignored. Ostriker \& McKee have given
asymptotic self-similarity solutions that incorporate this latter
complication, but unfortunately they are not sufficiently accurate for
our needs. The reason is that since neither energy nor momentum is
conserved in the regime before the shell becomes self-similar, there
is no accurate way to normalize the self-similar solution using the
initial data. 

Let $\rb$ and $\rd$ denote the average densities of baryonic and
non-baryonic matter in the universe. We will assume that all baryons
are in diffuse for early on, so that $\rb$ is also the density of the 
IGM. 
We will write $\rb = \Ob\rc$ and 
$\rd = \Od\rc$, where the critical density $\rc\equiv 3H^2/8\pi G$.
\noindent
We will use a three-phase model for the expanding bubbles:

\begin{itemize}

\item a dense, fairly cool spherical shell of outer radius $R$ and
thickness $R\delta$, containing a fraction $(1-\fm)$
of the total
baryonic mass enclosed,
 
\item uniform neutral ambient intergalactic medium (IGM) of density
$\rb+\rd$ and zero pressure outside,
 
\item a hot, thin, isothermal plasma of
pressure $p$ and temperature $T$ inside the shell.

\end{itemize}

\noindent
The shell is driven outwards by the pressure of the hot interior
plasma but slowed by the IGM and by gravity. The plasma is heated by
kinetic energy from supernova explosions and collisions with IGM
and cooled by bremsstrahlung and Compton drag against the cosmic
background radiation.
 
\subsection{The expanding shell}

We assume that the expanding shell sweeps up almost all baryonic IGM
that it encounters and loses only a small fraction of it through
evaporation into the interior, so that its total mass is
given by $m(t) = {4\over 3}\pi R(t)^3 (1-\fm)\rb$, where the
constant $\fm\ll 1$. Since $\dot{\rb}/\rb = -3H$ for any cosmological
model, we get  $${\md\over m} = \left(R^3 \rb\right)^{-1}
{d\over dt} \left(R^3 \rb\right) = 
3\left(\rdr-H\right)\hbox{ if }\rdr > H\hbox{, zero otherwise.}$$
(The shell will acquire new mass when it is expanding faster than
the Hubble flow, and will never lose mass.)
It turns out that the Hubble flow catches up with the shell
only as $t\to\infty$, so we will always have $\Rd>HR$ and $\md>0$.

When new mass is swept up,
it must be accelerated from the velocity $HR$ to $\Rd$, so the shell
experiences a net braking force $(\Rd-HR)\md$. 
The interior
pressure $p$ drives the shell outward with a force 
$pA = 4\pi R^2p = 3mp/\rb R$ in the thin shell approximation 
$\delta, \fm \ll 1$.  Finally there is a gravitational braking force,
which in the thin-shell approximation (Ostriker \& McKee 1988)
gives the deceleration ${4 \over 3} \pi G R(\rd + {1\over 2}\rb)$.
Adding these three force
terms, the radial equation of motion becomes
\beq{RadialEq}
\Rdd = {8\pi pG \over \Ob H^2R} - 
{3 \over R}\left(\Rd-HR\right)^2 - 
\left(\Od+{1\over 2}\Ob\right){H^2 R \over 2}.
\eeq

\subsection{The interior plasma}

The equation of state for the plasma in the hot interior gives the
thermal energy 
\beq{EqOfState}
\Et = {3 \over 2} pV = 2\pi pR^3,
\eeq
and energy conservation for the interior yields
\beq{EconsEq}
\dot \Et = L - pdV/dt = L-4\pi p R^2\Rd,
\eeq
where the luminosity $L$ incorporates all sources of heating
and cooling of the plasma. We will consider five contributions to
$L$ and write  
$$L = \Lsn-\Lc-\Lb-\Li+\Lh,$$
where $\Lsn$ is the energy injection from supernova explosions, 
$\Lc$ the cooling by Compton drag against the CBR,
$\Lb$ the cooling by bremsstrahlung,
$\Li$ the cooling by ionization of neutral IGM and  
$\Lh$ the heating from collisions between the shell and the IGM.

In stellar burning from zero to solar
metallicity, the mass fraction $0.02 \times 0.007$ is released,
mostly as radiation. Due to low cross-sections, only a negligible
fraction of this radiation will contribute towards heating the gas,
so we will only be interested in the energy that
is released in kinetic form. From empirical observations of active
galactic winds (Heckman 1990) about $2\%$ of the total luminosity from
a galaxy is mechanical. Another empirical observation is that for a
solar neighborhood initial stellar mass function, one has 
roughly one supernova for every $150\Ms$ of baryons that form
stars, with a  typical kinetic energy output of
$10^{51}$ ergs per explosion. Both of these observations lead to the
same estimate $$\Lsn = {\fsn M_b c^2\over\tb} \theta(\tb-t) \approx
1.2\Ls{M_b\over\Ms}\theta(\tb-t),$$  where the efficiency
$\fsn\aet{4}{-6}$ and where we have assumed that the  energy is
released at a constant rate during a  period $\tb\aet{5}{7}$ years.

Now let us examine cooling. 
The interior baryon density is $\ri = \rb\fm/(1-\delta)^3$
whereas the shell density is $\rs =
\rb(1-\fm)/(1-(1-\delta)^3) \approx \rb/3\delta$ if $\fm, \delta
\ll 1$. 
Compton drag against the
microwave background radiation causes energy loss at a rate
(Kompaneets 1957) 
\beq{LcompEq}
\Lc = {4\pi^2\over
15}\left(\st c n_e\right) \left({kT_e\over m_ec^2}\right) 
\left({k\Tg\over\hbar c}\right)^4\hbar c V,
\eeq
where $\st$ is the Thomson cross section, $V={4\over 3}\pi R^3$, 
and $T_e = T$, the temperature of the interior plasma, which is given
by
$$E_t = \left({3\over 2}+{3\over 2}\right)kT{\fm\rb\over m_p}V.$$
(We will assume almost complete ionization and low metallicity, so
that  $n_e\approx \fm n_b$.) Using \eq{EqOfState}, we see that
Compton drag causes cooling on a timescale 
$${\Et\over\Lc} = 
 {45\over 4\pi^2}\left({\hbar c\over k\Tg}\right)^4{m_e\over\st\hbar}
\aet{2}{12}\,\hbox{years}\times(1+z)^{-4},$$
that is, it becomes important only at high redshifts. 
It turns out that $\Lb\ll\Lc$ in our regime of interest, so we will
simply make the approximation $\Lb\approx 0$.
Assuming that the ambient IGM is completely neutral, the power
required to ionize the hydrogen entering the interior is
simply 
$$\Li = \fm n_b\rydberg\times 4\pi R^2\left[\Rd-HR\right],$$
where $\rydberg \approx 13.6\,eV$.

The equation of motion\eqnum{RadialEq} assumes that the
collisions between the expanding shell and the ambient IGM are
perfectly inelastic. The kinetic energy dissipated has one of three
fates: It may 

\begin{description}

\item[(a)] radiate away in shock cooling processes,
 
\item[(b)] ionize the swept up IGM, or
 
\item[(c)] heat the shell and by conduction heat the
interior plasma.
 
\end{description}

\noindent
Let $\fcoll$ denote the fraction that is reinjected
into the interior plasma through processes (b) and (c). 
This is one of the major uncertainties of the model.
Now a straightforward kinematic calculation of the kinetic
energy loss per unit time gives 
$$\Lh = \fcoll {3m\over 2R}\left(\Rd-HR\right)^3.$$
Making accurate estimates of $\fcoll$ is difficult, so we simply
use the two extreme cases $\fcoll = 0$ and $\fcoll = 1$ in the
simulations. Perhaps surprisingly, the results will be seen to be
relatively independent of the choice of $\fcoll$.

\subsection{Solutions to the equations}

Combining\eqnum{EqOfState} and\eqnum{EconsEq} leaves us
with
\beq{pdotEq}
\pd = {L\over 2\pi R^3} - 5 {\Rd\over R}p.
\eeq
The system of equations\eqnum{RadialEq} and\eqnum{pdotEq}
reduces to that derived by Weaver {\it et al.} (1977) in the
special case where $L(t)$ is constant and the expansion of
the universe is ignored.

Let us define dimensionless variables as follows:

\smallskip
{
\raggedright
\noindent
\baselineskip20pt
\tabskip = 1em
\halign{\hglue0.3cm$#\equiv$&$#, $&$#_*\equiv$&$#$&$#$\hfill\cr
\t&t/t_*&t&{2\over 3}H_0^{-1}(1+z_*)^{-3/2}&\cr
\h&H/H_*&H&{2\over 3}t_*^{-1}&\cr
\l&L/L_*&L&\fsn M_b c^2\tb^{-1}&\aet{1.2}{5}\Ls\times M_5\cr 
\e&E/E_*&E&L_*\tb&\aet{7.2}{53}\,\ergs\times M_5\cr 
\r&R/R_*&R&L_*^{1/5}G^{1/5}t_*&\approx 0.13\,\Mpc \times
 h^{-1}(1+z_*)^{-3/2}M_5^{1/5}\cr 
\p&p/p_*&p&L_*^{2/5}G^{-3/5}t_*^{-2}&\aet{1.4}{-16}\,Pa
\times h^2(1+z_*)^{3}M_5^{2/5}\cr}
}
\smallskip\noindent
Here we have taken $h=0.5$ and defined $M_5 \equiv M_b/10^5\,\Ms$.
If $\Omega\equiv\Ob+\Od$=1, then $t_*$ is the age of the
universe at the redshift $z_*$ when the shell begins its expansion,
i.e. the Big Bang occurred at $\t=-1$ and the shell starts expanding
at $\t=0$. For this simple case, we have $\h =
(1+\t)^{-1}=(1+z_*)^{-3/2}(1+z)^{3/2}$.
\goodbreak

Equations\eqnum{RadialEq} and\eqnum{pdotEq} now become
\beq{MainEq}
\cases{
\r''(\t) =&${18\pi\over\Ob}\h(\t)^{-2}{\p(\t)\over \r(\t)} 
-3\left(1-{2\over 3}{\h(\t)\r(\t)\over\r'(\t)}\right)^2
{\r'(\t)^2\over \r(\t)}
-\left({2\over 9}\Od+{1\over 9}\Ob\right) \h(\t)^2\r(\t)$\crr 
\p'(\t) =&${\l(\t)\over 2\pi \r(\t)^3} - 
5{\r'(\t)\over \r(\t)}\p(\t)$
}
\eeq
Here $\l=\ls-\lc-\li+\lh$, where 
\smallskip
{
\baselineskip20pt
\tabskip = 1em
\halign{\hglue3cm$#$&$#$&$#$\hfill\cr
\ls&=&\theta(\tb-t_*\t),\cr
\lc&\approx&0.017 h^{-1}(1+z_*)^{-3/2}(1+z)^4\r^3\p,\cr
\li&\approx&2.2\fm\Ob M_5^{-2/5}\times\h^2\r^2\hubpar\hbox{, and}\cr 
\lh&=&{1\over
3}\fcoll\Ob\times
  \left(\h\r\right)^2\hubpar^3.\cr
}}
\smallskip\noindent
In computing $\lc$, we have taken $T_{\gamma 0} =
2.74K$. The interior temperature, the thermal energy and the kinetic
energy are given by  
$$T \aet {4.5}{5}\,K\times 
 {M_5^{2/5}\over\fm\Ob}{\p(\t)\over\h(\t)^2},$$
$$\et = {2\pi\over\taub}\r^3\p,$$
$$\ek = {m\Rd^2/2 \over L_*\tb} = {\Ob\over 9\taub}\h^2\r^3\r'^2.$$

The solution to the system\eqnum{MainEq} evolves through three
qualitatively different regimes: $\t\ll 1$, $\t\approx 1$ and $\t\gg
1$.

\begin{description}

\item[a)] 
In the limit of small times $\t\ll 1$, gravity and Hubble
flow are negligible and we obtain the asymptotic power
law solution 
$$\r(\t) = a\t^{3/5},\quad \p(\t) = b\t^{-4/5}$$
\beq{SmallTimeSolEq}
\hbox{where } 
a\equiv\left({375/\Ob\over 77-27\fcoll}\right)^{1/5}
\quad\hbox{and}\quad
b\equiv{7\Ob\over 150\pi}a^2,
\eeq
as may be verified by direct
substitution. This solution reduces to that found by Weaver {\it et al.}
in the special case $\fcoll = 0$. 
Since the total energy injected is
simply $\ein=\t/\taub$, this gives  $${\et\over\ein} = {35\over
77-27\fcoll}\quad\hbox{and}\quad
{\ek\over\ein} = {15\over 77-27\fcoll}$$ for small $\t$.
Hence even though $\et+\ek=\ein$ only for the most optimistic
case $\fcoll = 1$, we see that no more than 
$1-{35+15\over 77}\approx 25\%$ of the injected energy is lost as
radiation even in the worst case $\fcoll=0$.

\item[b)] 
The behavior in the intermediate regime is a complicated interplay
between several different effects:

\begin{enumerate}

\item After approximately $5\times10^7$ years, the supernova explosions
cease, which slows the expansion. In this pressure-driven snowplow
phase, we would asymptotically have $R\propto t^{2/7}$, $t^{4/13}$,
$t^{1/3}$, $t^{4/11}$ or $t^{2/5}$ if there were no gravity, no Hubble
flow and no cooling with $\fcoll=0$, ${1\over 8}$, ${1\over 3}$,
${5\over 8}$ or $1$, respectively.
 
\item Cooling (and $pdV$) work reduces the pressure and the
thermal energy to virtually zero, which slows the expansion.
With zero pressure, we would approach the momentum-conserving
snowplow solution $R\propto t^{1/4}$ if there were no gravity and no Hubble
flow.
 
\item The density of the IGM drops and the IGM already has an outward
Hubble velocity before it gets swept up, which
boosts the expansion and adds kinetic energy to the shell.
 
\item Gravity slows the expansion.
 
\item 
Dark
matter that has been accelerated outward by the shell catches up with
it again and speeds up the expansion.
(This last effect has been neglected in the equations above, since it
generally happens too late to be of importance for our purposes.)

\end{enumerate}
 
\item[c)]
As $t\to\infty$, the shell gets
frozen into the Hubble flow, {\ie} $R\propto t^{2/3}$ if $\Omega=1$. An
approximate analytic solution for $\t\gg 1$ is given by Ostriker\&
McKee (1988), but since neither energy nor momentum is conserved in the
intermediate regime,  there is no simple way to connect this solution
with the short-time solution above. 

\end{description}

\begin{figure}[phbt]
\centerline{\epsfxsize=14cm\epsfbox{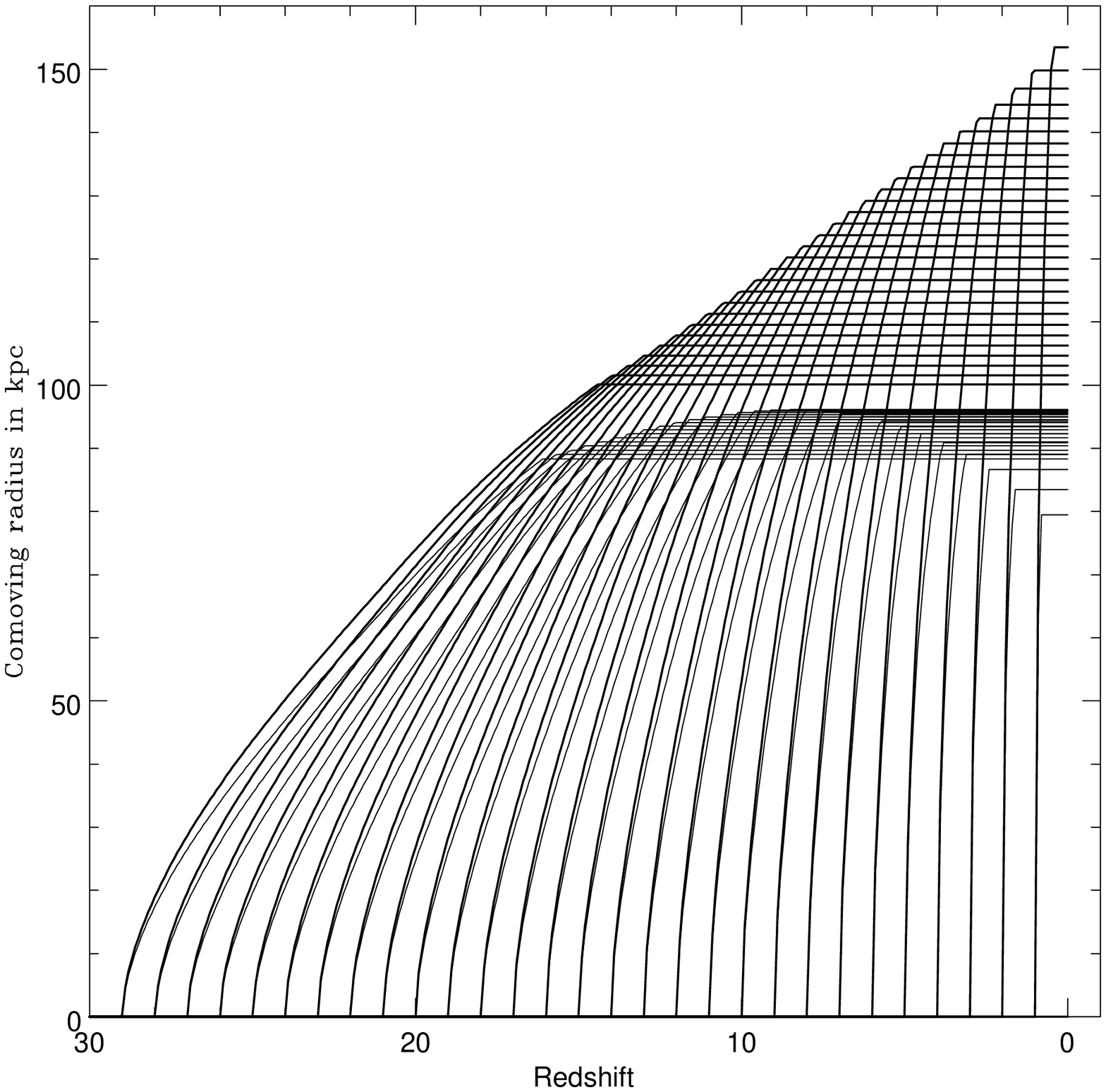}}
\caption{Comoving radius of expanding shell.}
\label{gpfig1}
The comoving shell radius $(1+z)R$ is plotted for galaxies
of total mass $2\times 10^6\Ms$, forming at integer
redshifts from 1 to 29.
Here $\Omega=1$, $\Ob=0.06$, $h=0.5$, and $\fm=0.1$.
$\fcoll=1$ for the upper set of lines and $\fcoll=0$ for the 
lower set. R has been truncated when $T$ drops below 15,000 K, 
after which
newly swept up IGM fails to become ionized.
\end{figure}

\begin{figure}[phbt]
\centerline{\epsfxsize=14cm\epsfbox{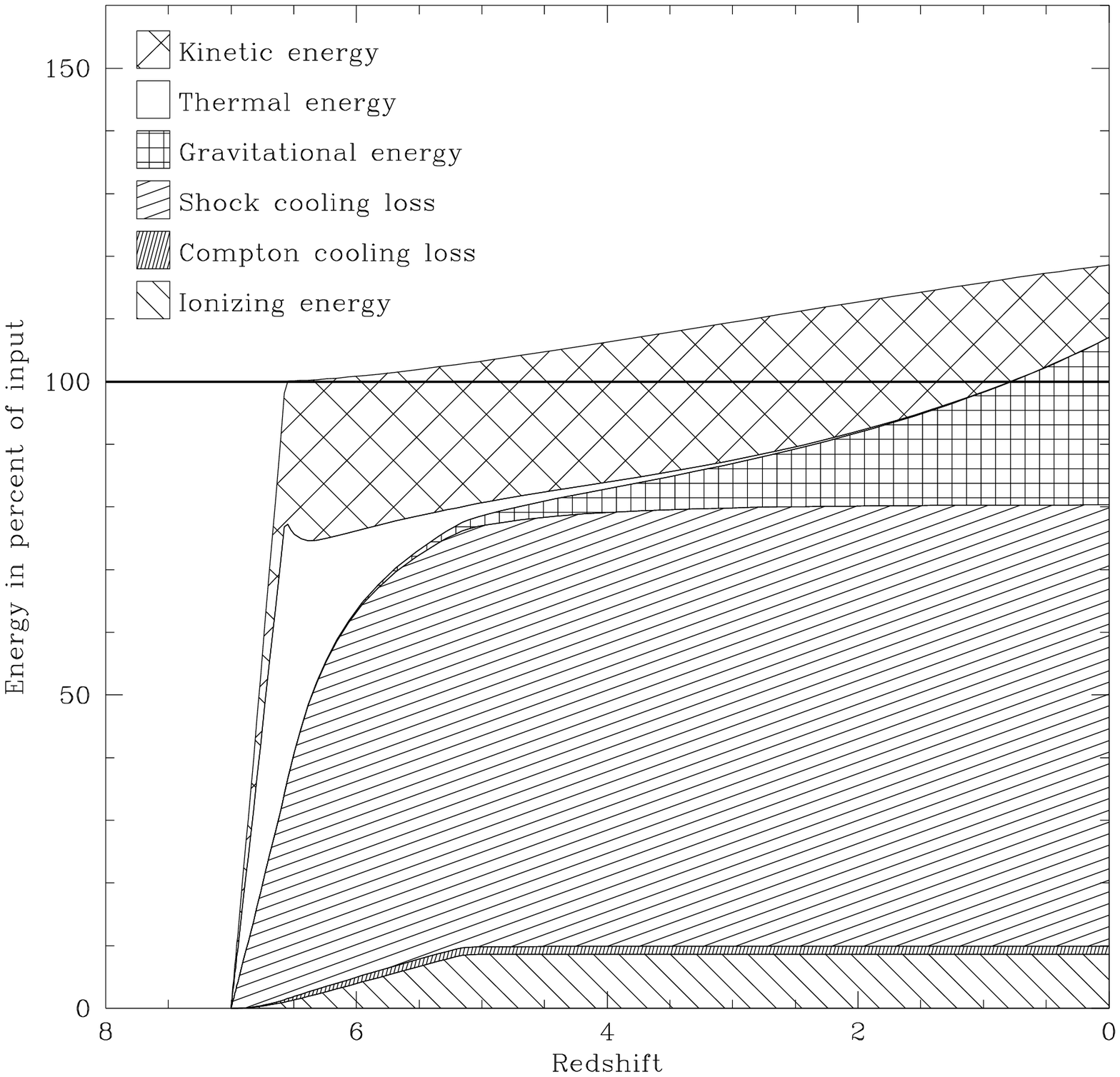}}
\caption{This and the two following 
figures show the
energy contents of an expanding bubble
as a function of redshift, for different choices of $\fcoll$
and $z_*$.
$\Omega=1$, $\Ob=0.06$, $h=0.5$ and $\fm=0.1$ for all three plots.
Figures~\ref{gpfig2a} and~\ref{gpfig2b}
illustrate the difference between $\fcoll=0$
and $\fcoll=1$ (there is no shock cooling loss in the second case).
Figure~\ref{gpfig2c}
has $\fcoll=0$ and illustrates that the Compton cooling
loss is larger at higher redshift.}
\label{gpfig2a}
\end{figure}

\begin{figure}[phbt]
\centerline{\epsfxsize=14cm\epsfbox{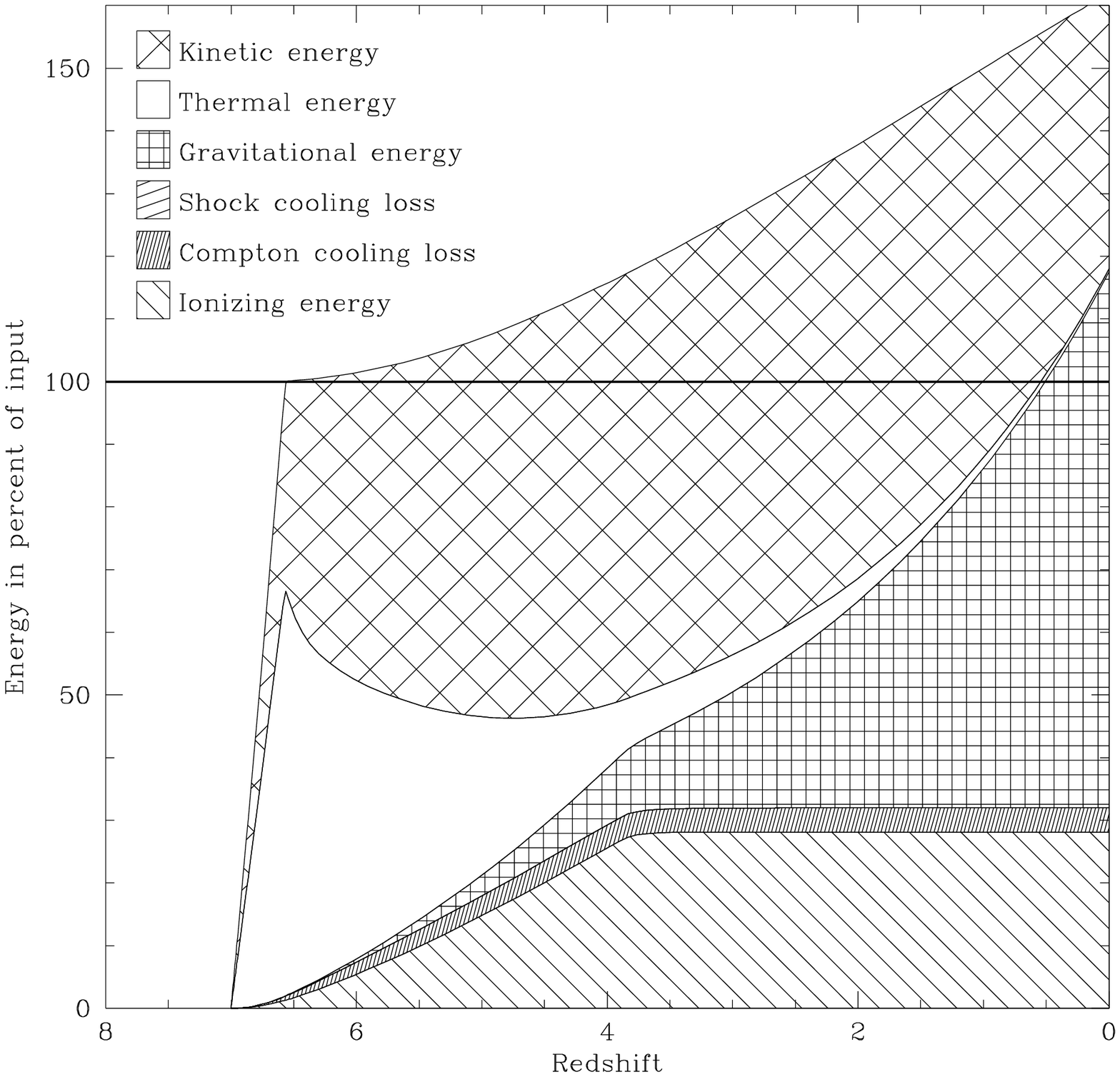}}
\caption{Energetics of expanding shell, example 2.}
\label{gpfig2b}
\end{figure}

\begin{figure}[phbt]
\centerline{\epsfxsize=14cm\epsfbox{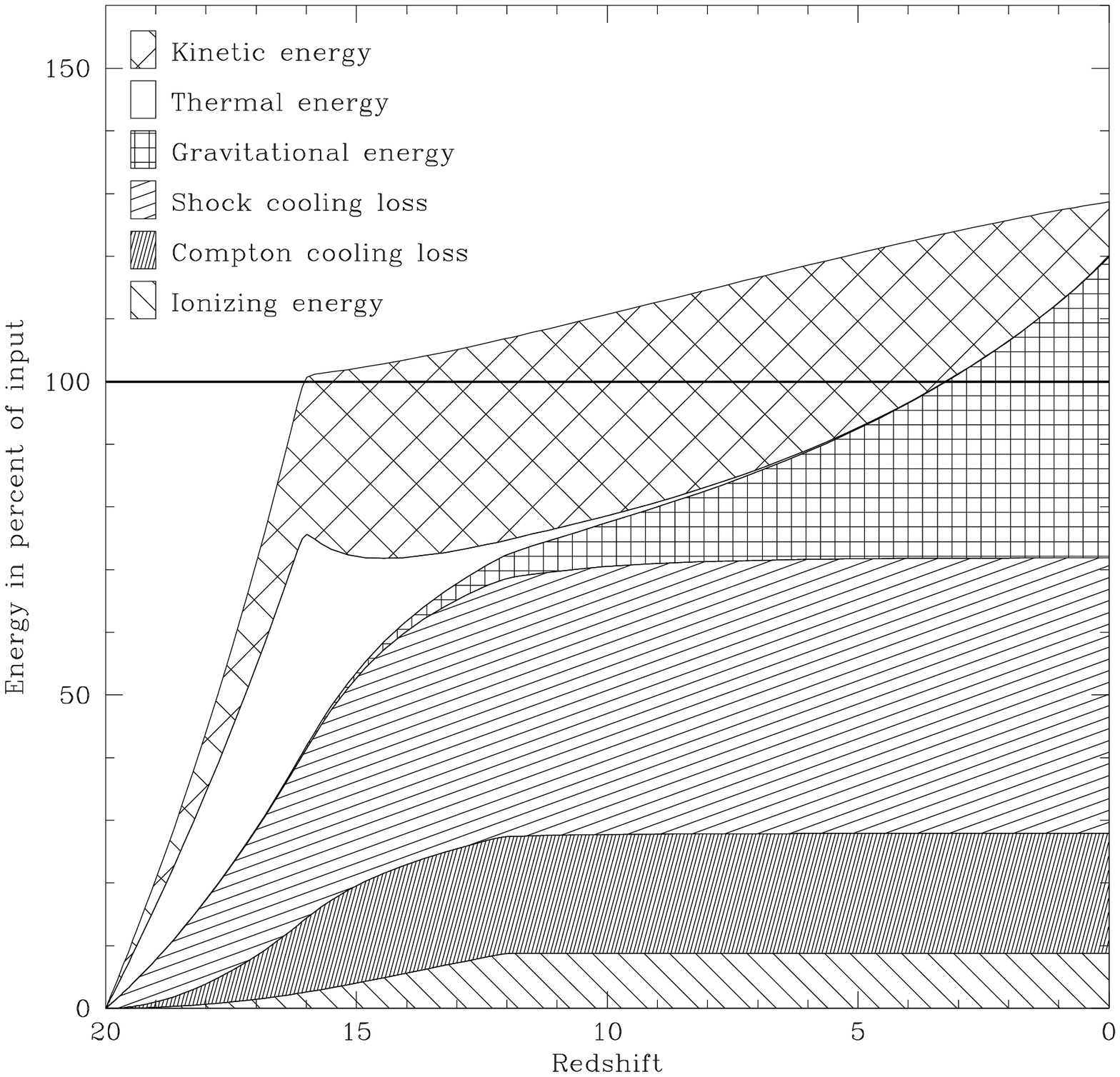}}
\caption{Energetics of expanding shell, example 3.}
\label{gpfig2c}
\end{figure}

\noindent
Numerical solutions for the comoving radius $(1+z)R$
are plotted in 
Figure~\ref{gpfig1} 
for different values of $z_*$ and $\fcoll$.
The asymptotic solution\eqnum{SmallTimeSolEq} has been used to
generate initial data at $\t=0.01$ for the numerical integration. 
In this Figure, 
we have truncated $R$ when the interior temperature
drops below 15,000K, after which newly swept up IGM fails to become
ionized.
Figures~\ref{gpfig2a}, \ref{gpfig2b} and~\ref{gpfig2c}  
show what becomes of the injected energy for
different parameter values.
Note that the relative fractions are approximately constant early on,
while the supernovae inject energy, in accordance with the
asymptotic solution\eqnum{SmallTimeSolEq}. The reason that the total energy
exceeds $100\%$ of the input is that the shell gobbles up
kinetic energy from swept-up IGM that already has an outward
Hubble velocity.

\section{Cosmological Consequences}
\label{gpsec3}

Once the expanding bubbles discussed in the previous section have
penetrated most of space, the IGM will presumably have a frothy
character on scales of a few 100 kpc, containing thick and fairly cool
shell fragments separated by large, hot, thin and ionized regions that
used to be bubble interiors.

In 
Section~\ref{gpsec3.1},
we calculate at what point the IGM becomes frothy,
more specifically what fraction of space is covered by expanding
shells at each $z$. 
In~\ref{gpsec3.2} 
we discuss the resulting enrichment of the
IGM with heavy elements. 
In~\ref{gpsec3.3} 
the thermal history of the IGM after
this epoch is treated.
Finally, in~\ref{gpsec3.4} 
the residual ionization is computed,
given this thermal history, and we discuss the circumstances under
which the Gunn-Peterson constraint is satisfied.

\subsection{IGM porosity}
\label{gpsec3.1}

Assuming the standard PS theory of structure
formation (Press \& Schechter 1974), the fraction of all mass that has
formed gravitationally bound objects of total (baryonic
and non-baryonic) mass greater than $M$ at redshift $z$ is
$$1-\erf\left[{\delta_c\over\sqrt{2}\sigma(M)}\right],$$  
where $\erf(x)\equiv 2\pi^{-1/2}\int_0^x e^{-u^2}du$ and 
$\sigma^2$ is the linearly extrapolated r.m.s. mass fluctuation in a
sphere of radius $r_0$. The latter is given by top-hat filtering of
the power spectrum as
\beq{FilterEq}
\sigma^2\equiv\left({\sigma_0\over 1+z}\right)^2 \propto
{1\over(1+z)^2}\izi
\left[{\sin kr_0\over(kr_0)^3} - {\cos kr_0\over(kr_0)^2}\right]^2
 P(k) dk,
\eeq
where $r_0$ is given by ${4\over 3}\pi r_0^3\rho = M$ and where
$P(k)$ is the power spectrum.
Although this approach has been criticized as too simplistic,
numerical simulations (Efstathiou {\it et al.} 1988; Efstathiou \&
Rees 1988; Carlberg \& Couchman 1989) have shown that it describes
the mass distribution of newly formed structures remarkably well.
Making the standard assumption of a Gaussian density field, Blanchard
{\it et al.} (1992) have argued that it is an accurate description at least in
the low mass limit. Since we are interested only in extremely low
masses such as $10^6\Ms$, it appears to suffice for our purposes.

We choose $\delta_c = 1.69$, which is the
linearly extrapolated overdensity at which a spherically symmetric
perturbation has collapsed into a virialized
object (Gott \& Rees 1975).
Letting $\fg$ denote the fraction of all baryons
in galaxies of mass greater than $M$ at $z$, this would imply that 
\beq{fgEq}
\fg \approx
1-\erf\left[{1.69(1+z)\over\sqrt{2}\sigma_0(M)}\right]
\eeq
if no other
forces than gravity were at work. However, it is commonly believed that
galaxies correspond only to such objects that are able to cool (and
fragment into stars) in a dynamical time or a Hubble time (Binney 1977;
Rees \& Ostriker 1977; Silk 1977; White \& Rees 1978). Hence the
above value of $\fg$ should be interpreted only as an upper limit.

A common assumption is that the first galaxies to form have a
total (baryonic and dark) mass $M_c\approx 10^6\Ms$, roughly the Jeans
mass at recombination.  Blanchard {\it et al.} (1992) examine the interplay
between cooling and gravitational collapse in considerable detail, and
conclude that the first galaxies to form have masses in the range
$10^7\Ms$ to  $10^8\Ms$, their redshift distribution still being given
by \eq{fgEq}. To keep things simple we will assume that all
early galaxies have the same mass $M_c$ and compare the results for
$M_c = 2\times 10^6\Ms, 10^8\Ms$ and $10^{11}\Ms$. 

Let $R(z;z_*)$ denote the radius of a shell at $z$ that was
created at $z_*$ by a galaxy of baryonic mass $M_b = \Ob M_c$ as in 
Section~\ref{gpsec2}.
Then the {\it naive filling factor}, the
total bubble volume per unit volume of the universe, is 
\beq{ffEq}
\ff(z) = \int_z^{\infty} 
{4\over 3}\pi R(z;z_*)^3 {\rb\over M_b}
{df_g(z_*)\over d(-z_*)} \,dz_*
= \phi_*(1+z)^3\int_z^{\infty} 
{r(z;z_*)^3\over (1+z_*)^{9/2}}
\,{df_g(z_*)\over d(-z_*)} \,dz_*,
\eeq
where
$$\phi_*\approx 1600h^{-1}M_5^{-2/5}(\Ob/0.06).$$
Clearly nothing prohibits $\ff$ from exceeding unity. This
means that nearby shells have encountered each other and that certain
volumes are being counted more than once. If the locations of the
bubbles are uncorrelated, then the fraction of the universe that will
be in a bubble, the {\it porosity}, is given by  
$$\P \equiv 1-e^{-\phi}.$$

If the early
galaxies are clustered rather than Poisson-distributed, this value
is an overestimate. 
For an extreme (and very unrealistic) example, if they would always
come in clusters of size $n$ and the clusters would be much smaller
than the typical bubble size of 100 kpc, then it is easy to see
that $\P \approx 1-e^{-\phi/n}.$ For more realistic cases, simple
analytic expressions for $P$ are
generally out of reach. Since we expect the clustering to be quite
weak, we will use the Poisson assumption for simplicity.

\begin{figure}[phbt]
\centerline{\epsfxsize=11cm\epsfbox{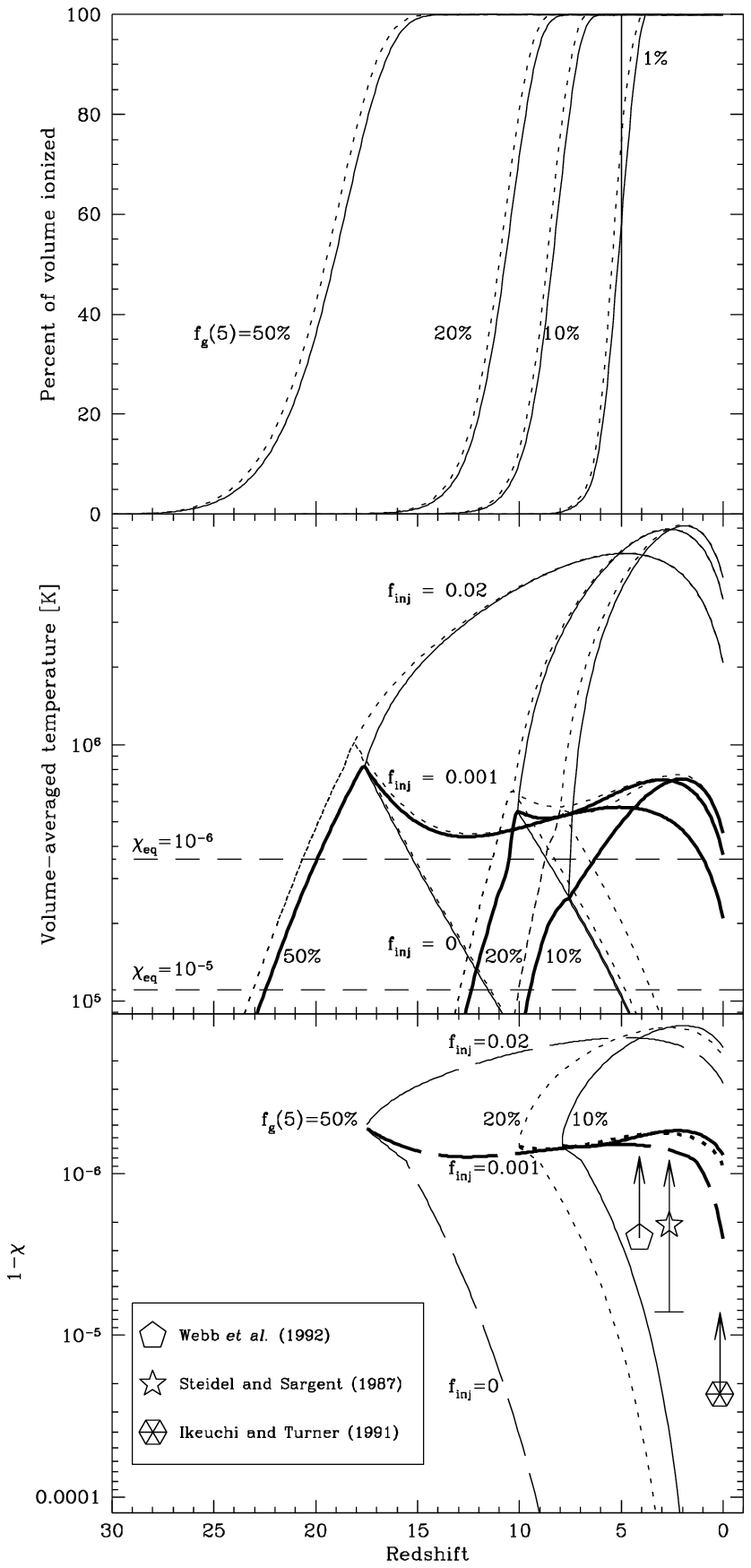}}
\caption{IGM evolution for $M_c=2\times 10^6\Ms$.}
\label{gpfig3a}
\end{figure}

\begin{figure}[phbt]
\centerline{\epsfxsize=11cm\epsfbox{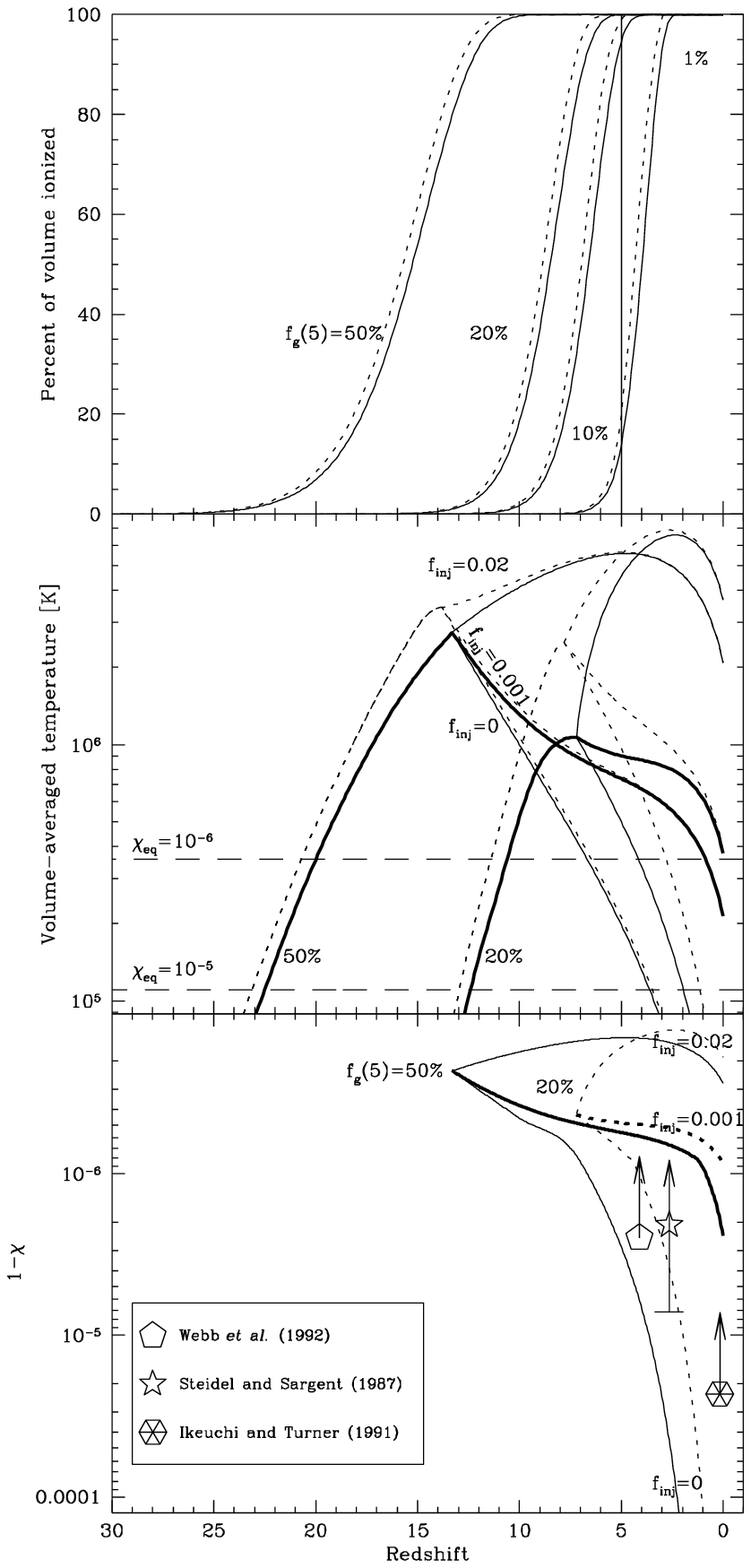}}
\caption{IGM evolution for $M_c=10^8\Ms$.}
\label{gpfig3b}
\end{figure}

The uppermost panels of 
Figures~\ref{gpfig3a} and~\ref{gpfig3b}  
contain $\P(z)$ for various
parameter values, calculated numerically from \eq{ffEq} using
the numerical solutions for $\r(z;z_*)$. 
It is seen that the lower mass in 
Figure~\ref{gpfig3a} 
($2\times 10^6\Ms$ 
versus $10^8\Ms$) gives higher filling factors, so that the expanding
shells fill almost $100\%$ of space by $z=5$ for three of the four 
choices of $f_g(5)$. In~\ref{gpfig3b}, we see that almost $20\%$ of 
the baryons must be in galaxies by $z=5$ to achieve this.
The greater efficiency of small galaxies 
is to be expected, since $\phi_*\propto M^{-2/5}$. Although
some parameters still yield the desired $P\approx 100\%$ by $z=5$ 
in Figure~\ref{gpfig3b},
using present-day masses like $M_c = 10^{11}\Ms$ fails dismally
(not plotted) for all choices of the other parameters. Roughly, the
largest $M_c$ that works is $10^8\Ms$

As can be seen, the dependence on
$\fcoll$ (dashed versus solid lines) is rather weak.

In order to calculate $\sigma_0(M_c)$ from the fluctuations observed
on larger scales today, we need detailed knowledge of the power
spectrum down to very small scales, something which is fraught with
considerable uncertainty. 
For this reason, we have chosen to label the curves by the more physical
parameter $f_g(5)$, the fraction of all baryons that have formed galaxies
by $z=5$. The four sets of curves correspond to fractions of $50\%$,
$20\%$, $10\%$ and $1\%$.
These percentages should be compared with observational
estimates of metallicity, as will be discussed in 
Section~\ref{gpsec3.2}.

The second column of 
Table~\ref{gptable1}  
contains the values $\sigma_0(M_c)$
necessary to obtain various values of $f_g(5)$, calculated by
inverting the error function in \eq{fgEq}. 
The last four columns contain the bias factors necessary to yield this
value of $\sigma_0(M_c)$ for two choices of power spectra
(CDM and n=0.7 tilted CDM) and two choices of cutoff
mass ($M_c = 2\times 10^6\Ms$ and $M_c =10^8\Ms$).
Thus $b =\gamma/\sigma_0(M_c)$, where we define
$\gamma$ to be the ratio between $\sigma$ at $M_c$ and 
$\sigma$ at $8h^{-1}\,\Mpc \equiv b^{-1}$.
Performing the integral\eqnum{FilterEq} numerically with the 
CDM transfer function given by Bardeen {\etal} 1986 (BBKS), 
$h=0.5$, $\Omega=1$, $\Omega_b \ll 1$ and an $n=1$ Harrison-Zel'dovich
initial spectrum gives $\gamma\approx 19.0$ for $M_c=2\times 10^6\Ms$ and
$\gamma\approx 13.6$ for $M_c=10^8\Ms$. Using the CDM transfer function
of Bond and Efstathiou (1984) instead gives $\gamma\approx 18.1$ and
$\gamma\approx 13.7$, respectively. The BBKS transfer function is 
more applicable here since it includes the logarithmic dependence that
becomes important for very low masses. The BBKS transfer function with a
tilted (n=0.7) primordial spectrum yields the significantly lower values 
$\gamma\approx 9.71$ and
$\gamma\approx 7.93$, respectively.

\begin{table}
$$
\begin{tabular}{|rrrrrr|}
\hline
$\fg(5)$&$\sigma_0(M_c)$&$b_{cdm,6}$&$b_{cdm,8}$&$b_{tilted,6}$&
$b_{tilted,8}$\\
\hline
1\%&3.94&4.8&3.5&2.5&2.0\\
10\%&6.18&3.1&2.2&1.6&1.3\\
20\%&7.92&2.4&1.7&1.2&1.0\\
50\%&15.06&1.3&0.9&0.6&0.5\\
\hline
\end{tabular}
$$
\caption{Correspondence between various ways of 
normalizing the power spectrum}
\label{gptable1}
\end{table}
\noindent
Basically, 
Table~\ref{gptable1}  
shows that any of our values of $f_g(5)$ become
consistent with a feasible bias factor for some choice of
power spectrum and cutoff mass. 
\bigskip

\subsection{IGM enrichment}
\label{gpsec3.2}

These values of $f_g(5)$ should be compared with observational
estimates of metallicity, since if the
stars in these early mini-galaxies produce the same fractions of heavy
elements as do conventional stars, then these percentages are directly
linked to the fraction of currently observed metals that were made
before $z=5$. Some of the enriched shells may be observable as quasar 
absorption line
systems, as intracluster gas, and, indirectly, as in the
metallicities of old disk and halo stars.

Observations of
iron abundances in intracluster gas by HEAO-1,
Exosat and Ginga ({\eg} Mushotzky 1984; 
Hughes {\it et al.} 1988; Edge 1989;
Hatsukade 1989) have shown that most clusters have abundances
between $25\%$ and $50\%$ of the solar value. Einstein observations
have showed the presence of a large variety of other heavy elements in
the intracluster gas (Lea {\it et al.} 1982; Rothenflug {\it et al.} 1984).
Most of this gas and some of these metals are believed to be
``primordial",  since the gas mass in clusters is typically several
times greater than the observed stellar mass in the cluster galaxies
(Blumenthal {\it et al.} 1984; David {\it et al.} 1990; Arnaud {\it et al.}
1991).

There are indications that the most of these heavy elements
may have been produced as recently as around $z=2-3$, and that the
metallicity in the halo gas of some $z\approx 3$ galaxies inferred
from QSO absorption line studies are as low as $0.1\%$ of the
solar value (Steidel 1990).
However, this and other
observations of extremely metal-poor objects 
(Pettini {\etal} 1990) does not necessarily rule out our
scenario, since it is highly uncertain whether all the hydrogen in the
swept-up IGM would get thoroughly mixed with the metal-rich supernova
ejecta.

\subsection{IGM temperature}
\label{gpsec3.3}

Let $T(z;z_*)$ denote the temperature of the interior of a bubble at
$z$ that was created at $z_*$ as in 
Section~\ref{gpsec2}.
Then the volume-averaged temperature of the IGM is 
\beqa{IGMtempEq}
\Tigm(z) 
&\equiv&\int_z^{\infty} 
{4\over 3}\pi R(z;z_*)^3 
T(z;z_*) {\rb\over M_b}
{df_g(z_*)\over d(-z_*)} \,dz_*\nonumber\\
&=& \phi_*(1+z)^3\int_z^{\infty} 
{r(z;z_*)^3\over (1+z_*)^{9/2}}\,
T(z;z_*)
{df_g(z_*)\over d(-z_*)} \,dz_*.
\eeqa
When $\phi$ becomes of order unity, the IGM swept up by the expanding
shells is no longer cold, neutral and homogeneous, so the
treatment in 
Section~\ref{gpsec2}
breaks down. The resulting temperatures will
be underestimated, since less thermal energy needs to be expended on
heating and ionization.  As can be seen in 
Figures~\ref{gpfig3a} and~\ref{gpfig3b},  
the
transition from $\phi\ll 1$, where the treatment in 
Section~\ref{gpsec2}
is valid, to $\phi\gg 1$, where the IGM becomes fairly uniform, is quite
rapid. Since $\Tigm$
defined above is proportional to the thermal energy per unit volume,
energy conservation leads us to assume that $\Tigm$ remains fairly
constant during this transition and therefore is a good estimate of
the bulk IGM temperature immediately afterwards.
From this time on, we will approximate the IGM outside the
scattered dense and cold shell remnants by a uniform isothermal 
plasma.
Applying equations\eqnum{EqOfState},\eqnum{EconsEq} and\eqnum{LcompEq}
to the IGM yields the
following equation for its thermal evolution: 
\beq{dTdzEq}
-{d\over dz} T_5 = -\left[{2\over 1+z} + A(1+z)^{3/2}\right] T_5
+ \linj,
\eeq
where the first term encompasses cooling from adiabatic expansion 
and the second term Compton cooling. $T_5\equiv \Tigm/10^5\,K$, 
$A\equiv
1.5(t_0/t_{comp})\approx 0.0042h^{-1}$ and $\linj\equiv t_0
L_{inj}/k\times10^5K$, where  $ L_{inj}$ is the power injected into
the IGM per proton from all heat sources combined.
The Compton cooling term is seen to increase with
redshift, equaling the adiabatic term at $z\approx 17h^{0.4} - 1$.

In the most pessimistic case of no reheating whatsoever, {\ie} for
$\linj=0$, \eq{dTdzEq} has the solution 
$$T \propto (1+z)^2 e^{0.4A(1+z)^{5/2}}.$$
A more optimistic assumption is that some fraction $\finj$ of the
total energy released from stellar burning in newly formed galaxies
continues to heat the IGM, i.e. 
$$\linj = f_{inj}\left(0.02\times0.007 m_p c^2\over
k\times10^5K\right) {df_g\over d(-z)} \approx 
\left({2.1\times 10^4\over \sigma_0}\right)
f_{inj}\exp\left(-{1\over 2}
\left[{1.69(1+z)\over\sigma_0(M_c)}\right]^2\right),$$ 
which would also incorporate other modes of energy injection such
as radiation. 
The middle panels in 
Figures~\ref{gpfig3a} and~\ref{gpfig3b}  
show the temperature resulting from
$\finj=0, 0.001$ and $0.02$ and different initial values. For $\P<0.8$,
$\Tigm$ has been calculated numerically from 
\eq{IGMtempEq} by
using the numerical solutions for $r(z;z_*)$. Then the value at the
redshift for which $\P=0.8$ has been used as initial data for
\eq{dTdzEq}. For comparison, two horizontal lines have been
added showing what temperatures would be required to obtain neutral
fractions of $10^{-6}$ and $10^{-5}$ in {\it equilibrium}, using
\eq{EquilibrionizationEq}.
Since the plasma is in fact out of equilibrium, these highly ionized
states can be maintained at much lower temperatures, as is seen in the
bottom plots.

\subsection{IGM ionization and the Gunn-Peterson effect}
\label{gpsec3.4}

Assuming that any neutral hydrogen in the remains of the shells
will have insufficient time to diffuse far into the hot
ionized regions that used to be shell interiors, we can treat the
latter as an isolated mixture of gas and plasma where the ionization
fraction $\chi$ evolves as  $$\dot\chi =
n\chi\left[(1-\chi)\Lion(T)-\chi\Lrec(T)\right],$$ 
and where the rates
for collisional ionization and recombination are given by
(Stebbins \& Silk 1986)
$$\Lion = \left<\sigma_{ci}v\right>\approx
7.2\,a_0^2\left({kT\over m_e}\right)^{1/2} e^{-\rydberg/kT},$$
$$\Lrec = \left<\sigma_{rec}v\right>\approx
{64\pi\over 3\sqrt{3\pi}}\alpha^4a_0^2c
\left({kT\over\rydberg}\right)^{-2/3},$$
where $a_0$ is the Bohr radius and $\alpha$ is the fine structure
constant. Changing the independent variable to redshift, the $\Omega=1$
case leaves us with
$$-{d\over dz}\ii = {3\over 2}{\fm\Ob\over\omdc}(1+z)^{1/2}\ii
\bigl[(1-\ii)\lion(\ii)-\ii\lrec(T)\bigr],$$
$$\lion(T) \equiv
n_{c0}t_0\Lion(T)\approx \left[5.7\times 10^4 h\Ob\right]
T_5^{1/2}e^{1.58/T_5}$$  
$$\lrec(T) \equiv n_{c0}t_0\Lrec(T)
\approx\left[0.16 h\Ob\right] T_5^{-2/3}.$$
For large enough $z$, the ionization fraction will adjust rapidly
enough to remain in a quasistatic equilibrium and hence be given by
$\dot\ii=0$, {\ie}
\beq{EquilibrionizationEq}
\ii =
\left[1+{\Lrec(T)\over\Lion(T)}\right]^{-1}\approx
\left[1+2.8\times10^{-6}T_5^{-7/6}e^{1.58/T_5}\right]^{-1}.
\eeq

The observed absence of a Gunn-Peterson trough in the spectra of
high-redshift quasars strongly constrains the density of neutral
hydrogen in the IGM. The most thorough study to date, involving eight
quasars (Steidel \& Sargent 1987), concluded that
$$\Omega_{H_I}(z=2.64)<(1.2\pm 3.1)\times 10^{-8}h_{50}^{-1}$$
if $\Omega=1$.
In our model this corresponds to
$(1-\ii) < (1.2\pm 3.1)\times 10^{-8}/(\fm\Ob)\aet{2}{-6}$ 
for $\Ob=0.06$ and $f_m$=0.1. Thus we
are helped not only by the IGM being ionized, but also by it being
diffuse. 
In a recent study of a single quasar, Webb {\etal} (1992) find the data
consistent with either $\Omega_{H_I}(z=4.1)=0$ or
$\Omega_{H_I}(z=4.1)=1.5\times
10^{-8}h_{50}^{-1}$, depending on model assumptions. We will
use the latter value as an upper limit.
Finally, recent Hubble Space Telescope spectroscopy of 3C 273 has been
used to infer that $\Omega_{H_I}(z = 0.158)<1.4\times
10^{-7}h_{50}^{-1}$. The constraints from these three studies are
plotted in 
Figures~\ref{gpfig3a} and~\ref{gpfig3b}  
together with the ionization levels
predicted by our scenario.

To achieve $\ii = 10^{-4}$, $10^{-5}$ and $10^{-6}$ in equilibrium
would by \eq{EquilibrionizationEq} require $T > 5.5\times
10^4\,K$, $T > 1.1\times 10^5\,K$ and $T > 3.6\times 10^5\,K$,
respectively.
As can be seen from the numerical solutions in the bottom panels of
Figures~\ref{gpfig3a} and~\ref{gpfig3b},  
the recombination rate is generally too slow for equilibrium
to be established, and the IGM remains almost completely ionized even when
$T\ll 15,000 K$ and equilibrium would have yielded $\ii\approx 0$. 
In both 3a and 3b, a very moderate reheating ($f_{inj} =
0.001$, heavy lines) is seen to suffice to satisfy the three
observational constraints. In the absence of any reheating whatsoever,
the only models that satisfy the constraints are those with very low
density ($\Ob=0.01$ or $\fm=0.01$). 

In summary, the only parameters that are strongly
constrained by the Gunn-Peterson test are $M_c$ and $\sigma_0(M_c)$.

\subsection{Other spectral constraints}

Let us estimate to what extent Compton cooling of the hot plasma
will distort the cosmic microwave background radiation (CBR). Since
for $T_e\gg\Tg$ the Comptonization y-parameter (Stebbins \& Silk
1986) 
$$y_C\equiv\int_t^{t_0}
{kT_e\over m_ec^2}n_e\st c\,dt$$ 
is linear in the plasma
energy density  
$\left({3\over 2}+{3\over 2}\right)kT_en_e$ at each fixed
time $(T_e = \Tigm)$, all that counts is the spatially averaged thermal
energy density at each redshift. 
Since the former is simply $\et(z;z_*)$
times the density of injected energy $f_g\fsn\Ob\rho c^2$,
the calculation reduces to mere energetics and we obtain
$$y_C = y_*\int_z^{\infty} 
{df_g(z_*)\over d(-z_*)} 
  \int_0^{z_*} \sqrt{1+z}\> \et(z;z_*) dz\,dz_*,$$
where
$$y_* \equiv{1\over 8} \fsn \Ob^2 {\st cH\over m_pG}
\aet{9}{-7}h\Ob^2.$$
The current observational upper limit on $y$ is  
$2.5\times 10^{-5}$ (Mather {\etal} 1994), so even if we 
take $f_g(0)$ as high as
$100\%$ and make a gross overestimate of the integral by making all
our galaxies as early as at $z_*=30$ and by
replacing $\et(z;z_*)$ by 
its upper bound $60\%$ for all $z$, $z_*$, our $y$ is
below the observational limit by three orders of magnitude for
$\Omega=1$. \smallskip

Now let us estimate the optical depth of the IGM. It has long been
known that reionization can cause a
spatial smoothing of the microwave background as CBR photons Thomson
scatter off of free electrons. Since $n_e  = \chi_{IGM}
n_b$, the optical depth for Thomson scattering, i.e. the number of
mean free paths that a CBR photon has traversed when it reaches our
detectors, is  
$$\tau_t = \int_{t_{rec}}^{t_0}\st \chi_{IGM} n_bc\, dt
= \tau_t^* \int_0^{z_{rec}}\sqrt{1+z} \chi_{IGM}
dz,$$ 
where
$$\tau_t^*\equiv {3\over 8\pi}\fm\Ob{H_0c\st\over
m_pG}\approx 0.07\Ob h.$$
Let us evaluate the integral by making the approximation that
$\chi_{IGM}$ increases abruptly from $0$ to $1$ at some redshift
$z_{ion}$. Then even for $z_{ion}$ as high as 30, 
$\tau_t \approx 7.9 h\Ob\fm\approx 0.02 \ll 1$ for our fiducial
parameter values $h=0.5, \Ob=0.06$ and $\fm=0.1$, so the probability
that a given CBR photon is never scattered at all is 
$e^{-0.02}\approx 98\%$. Hence this scenario for late reionization
will have only a very marginal smoothing effect on the CBR.
If the shells are totally ionized as well, then the factor $\fm$
disappears from the expressions above which helps only slightly.
Then $z_{ion}=15$ would imply that $8\%$ of the CBR
would be spatially smoothed on scales of a few degrees.

\section{Discussion}

We have calculated the effects of supernova driven winds from
early galaxies assuming a Press-Schechter model of galaxy formation
and a CDM power spectrum. The calculations have shown that
reionization by such winds can indeed explain the observed
absence of a Gunn-Peterson effect if a number of conditions are
satisfied:

\begin{enumerate}

\item The masses of the first generation of galaxies must be very small,
not greater than about $10^8\Ms$.
 
\item There is enough power on these small scales to get at least $10\%$
of the baryons in galaxies by $z=5$.
 
\item Except for the case where $\Ob$ is as low as $0.01$, there must be
some reheating of the IGM after $z=5$ to prevent the
IGM from recombining beyond allowed levels.
 
\item The commonly used thin-shell approximation
for
expanding bubbles must remain valid over cosmological timescales, with
the mass fraction in the interior remaining much less than
unity.

\end{enumerate}
 
\smallskip

Whether 1) is satisfied or not depends crucially on the model for
structure formation. This scenario is consistent with a pure CDM model
and some low-bias tilted CDM models, but not with top-down models like
pure HDM.

\smallskip

Observations of nearly solar abundances of heavy elements in
intracluster gas have given some support for 2),
which is roughly  equivalent to requiring that at least $10\%$ of the
heavy elements in the universe be made before $z=5$ (or
whenever $\phi\gg 1$). As discussed in 
Section~\ref{gpsec3.2},
the observations of some extremely metal-poor objects in QSO absorption
line studies do not necessarily rule out our scenario, since it is
highly uncertain whether all the hydrogen in the swept-up IGM would get
thoroughly mixed with the metal-rich supernova ejecta. 
The fact that large numbers of mini-galaxies are not seen today need not
be a problem either. Possible explanations for this
range from mechanisms for physically destroying them 
(Dekel \& Silk 1986, for instance) to the fact that 
the faint end of the luminosity function is still so poorly known
that old dwarf galaxies in the field may be too faint to see
by the present epoch (Binggeli {\etal} 1988).
\smallskip

To violate 3), the actual reheating would have to be extremely small.
A current IGM temperature between $10^4K$ and $10^5K$ suffices,
depending on other parameter values, since the low density IGM never has
time to reach its equilibrium ionization.
\smallskip

The thin-shell approximation 4) is obviously a weak point in the
analysis, because of the simplistic treatment of the dense shell and
its interface with the interior bubble. For instance, could the shell
cool and fragment due to gravitational instability before it
collides with other shells?
An approximate analytic model for such instability has been
provided by Ostriker \& Cowie (1981). Their criterion is that
instability sets in when $\Xi>1$, where
$$\Xi\equiv {2G\rho_{shell}R^2\over \Rd v_s}$$
and the sound speed $v_s = \sqrt{5 kT/3 m_p}$. In terms of
our dimensionless variables, this becomes
$$\Xi\approx 0.011 \times M_5^{1/5} T_5^{-1/2} 
\left({\Ob\over 0.06}\right)\left({\delta\over 0.1}\right)^{-1}
\left({1+z\over 1+z_*}\right)^3{r^2\over r'},$$ 
which indicates that with our standard parameter values,
gravitational instability does not pose problems even with fairly low
shell temperatures.
The reason that the shell density is not limited to four times the 
ambient IGM density is that 
the jump condition is not adiabatic, due mainly to effective Compton
cooling at the high redshifts under consideration.

After the critical $z$ (typically between 20
and 5) at which the expanding shells have collided with
neighbors and occupied most of space, the IGM is ``frothy" on
scales around 100 kpc, with dense cool shell remnants scattered
in a hot thin and fairly uniform plasma. Since the dark matter
distribution is left almost unaffected by the expanding bubbles,
formation of larger structures such as the galaxies we observe today 
should remain fairly unaffected as far as concerns gravitational
instability. There is indirect influence, however: the
ubiquitous metals created by the early mini-galaxies would enhance the
ability of the IGM to cool, which as mentioned in 
Section~\ref{gpsec3.1}
is commonly believed to be crucial for galaxy formation. 

Blanchard {\it et al.} (1992) argue that if the IGM has a temperature
higher than the virial temperature of a dark halo, pressure support
will prevent it from falling into the potential well and thus stop
it from forming a luminous galaxy. The virial temperature they
estimate for an object of mass $M$ formed at a redshift $z$ is
approximately
$$T_{vir} \aet{5.7}{5}K\,\left({M\over 10^{12}\Ms}\right)^{2/3}(1+z)$$
for $h=0.5$, so requiring $T_{vir}>T_{IGM}$ for say $T_{IGM}=10^6K$ at
$z=5$ would give a minimum galaxy mass of about $10^{11}\Ms$.
Such arguments indicate that the IGM reheating of our scenario might
produce a ``mass desert" between the earliest mini-galaxies and the
galaxies we see today: 
The first generation of galaxies, mini-galaxies
with masses of perhaps $10^6$ or $10^8 \Ms$, would keep forming until
their expanding bubbles had occupied most of space and altered the
bulk properties of the IGM. After that, formation of galaxies much
smaller than than those of today would be suppressed, since the IGM
would be too hot. 
Eventually, as the IGM cools by adiabatic expansion, a
progressively larger fraction of the IGM can be accreted by dark
matter potential wells. 
Indeed, even with the volume averaged IGM temperature remaining hot due
to some form of reheating,  cooling flows in deep potential wells, in
particular galaxy clusters, would not be suppressed. Late formation of
galaxies is therefore possible.

Pressure balance between the shell and the interior
during the expansion would give the ration
$T_{shell}/T_{interior} = \rho_{b,interior}/\rho_{b,shell}\approx
3\delta\fm\approx 0.03$ for $\delta=\fm=0.1$, so the 
shell fragments would expected to contain non-negligible fractions of
neutral hydrogen and thus absorb some Lyman-alpha. 
A typical shell radius is about 100 kpc for $M_c=10^6\Ms$, 
a size comparable to that of the clouds of the Lyman-alpha forest.
As to the number density of Lyman-alpha
clouds, the observed velocity separations greatly exceed those we
would expect if all shell fragments were to be identified with Lyman
alpha clouds. Thus the majority of these fragments must have been
destroyed by some other process. There are a number of ways in which this
could occur, for instance through photoionization by UV flux from the
parent galaxy or by collapse to form
other dwarf galaxies. The resulting numbers resemble the abundance of
minihalos in an alternative explanation of the Lyman-alpha forest
(Rees 1986). Strong evolution, in the sense of an increasing
cloud abundance with decreasing redshift, is expected to 
occur as cooling becomes effective.

\bigskip
The authors would like to thank Alain Blanchard for
discussions on the subject of the paper and David
Schlegel, Douglas Scott and Charles Steidel for many useful comments. 
This research has been supported in part by a grant from the NSF.

\section{REFERENCES}

\rf Arnaud, M., Rothenflug, R., Boulade, O., Vigroux, L. \&
Vangioni-Flam, E. 1992;Astr. Ap.;254;49

\rf Arons, J. \& McCray, R. 1970;Ap. Letters;5;123
 
\rf Bardeen, J. M., Bond, J. R., Kaiser, N. \& Szalay, A. S.
1986;ApJ;304;15
 
\rf Bergeron, J. \& Salpeter, E. E. 1970;Ap. Letters;7;115

\rf Binney, J. 1977;ApJ;215;483

\rf Binggeli, B., Sandage, A. \& Tammann, G. A. 
1988;Ann. Rev. Astr. Ap.;26;509 
 
\rf Blanchard, A., Valls-Gabaud, D. \& 
Mamon, G. A. 1992;Astr. Ap.;264;365

\rf Blumenthal, G. R., Faber S. M., Primack, J. R., \&
Rees, M. J.1984;Nature;311;517
 
\rf Bond, J. R. \& Szalay, A. S. 1983;ApJ;274;443 

\rf Bond, J. R. \& Efstathiou, G. 1984;ApJ (Letters);285;L45 

\rf Bruhweiler, F. C., Gull, T. R., Kafatos, M., \&
Sofia, S. 1980;ApJ (Letters);238;L27
 
\rf Carlberg, R. G. \& Couchman, H. M. P. 1989;ApJ;340;47
 
\rf Cioffi, D. F., McKee, C. F., \&
Bertschinger, E. 1988;ApJ;334;252

\rf Couchman, H. M. P. \& Rees, M. 1988;MNRAS;221;53
 
\rf Cox, D. P. \& Smith, B. W. 1974;ApJ (Letters);189;L105
 
\rf David, L. P., Arnaud, K. A., Forman, W., \&
Jones, C. 1990;ApJ;356;32
 
\rf David, L. P., Forman, W., \& Jones, C. 1991;ApJ;369;121
 
\rf Davis, M., Summers, F. J. \& Schlegel, D. 1992;Nature;359;393

\rf Dekel, A. \& Silk, J. 1986;ApJ;303;39

\rf Donahue, M. \& Shull, M. 1987;ApJ (Letters);323;L13
 
Edge, A. C. 1989, Ph.D. Thesis, University of Leicester.
 
\rf Efstathiou, G. \& Rees, M. J. 1988;MNRAS;230;5P
 
\rf Efstathiou, G., Frenk, C. S., White, S. D. M., \&
Davis, M. 1985;ApJ;57;241
 
\rf Efstathiou, G., Frenk, C. S., White, S. D. M., \&
Davis, M. 1988;MNRAS;235;715
 
\rf Feynman, R. P. 1939;Phys. Rev.;56;340

\rf Gott, J. R. \& Rees, M. J. 1975;Astr. Ap.;45;365 
 
\rf Gunn, J. E. \& Peterson, B. A. 1965;ApJ;142;1633
 
Hatsukade, I. 1989, Thesis, Osaka University.
 
\rf Heckman, T. M., Armus, L., \& 
Miley, G. K. 1990;ApJ Suppl.;74;833

\rf Holtzman, J. A. 1989;ApJ Suppl.;71;1

\rf Hughes, J. P., Yamashita, K., Okumura, Y.,
Tsunemi, H., \& Matsuoka, M. 1988;ApJ;327;615
 
\rf Ikeuchi, S. \& Ostriker, J. P. 1986;ApJ;301;522
 
\rf Lea, S. M., Mushotzky, R., \& Holt, S. 1982;ApJ;262;24.
 
\rf McCray, R. \& Kafatos, M. 1987;ApJ;317;190
 
\rf McCray, R. \& Snow, T. P. Jr. 1979;Ann. Rev. Astr. Ap.;17;213
 
\rf McKee, C. F. \& Ostriker, J. P. 1977;ApJ;218;148
 
\rf Miralda-Escude, J. \& Ostriker, J. P. 1990;ApJ;350;1
 
\rf Mushotzky, R. F. 1984;Physica Scripta;T7;157
 
\rf Ostriker, J. P. \& Cowie, C. F. 1981;ApJ (Letters);243;L115
 
\rf Ostriker, J. P. \& McKee, C. F. 1988;Rev. Mod. Phys.;60;1

\rf Peacock, J. A. \& Heavens, A. F. 1990;MNRAS;243;133
 
\rf Pettini, M., Boksenberg, A., \& Hunstead, R. 1990;ApJ;348;48
 
\rf Press, W. H. \& Schechter, P. 1974;ApJ;187;425
 
\rf Rees, M. J. \& Ostriker, J. P. 1977;MNRAS;179;541

\rf Rees, M. J.1986;MNRAS;218;25P
 
\rf Rothenflug, R. L., Vigroux, R., Mushotzky, R.,
\& Holt, S. 1984;ApJ;279;53

\rf Schwarz, J., Ostriker, J. P. \& Yahil, A. 1975; ApJ;202;1

Sedov, L. I. 1959; {Similarity and Dimensional Methods in
Mechanics} (Academic, New York).
 
\rf Shafi, Q. \& Stecker, F. W. 1984;Phys. Rev. Lett.;53;1292
 
\rf Shapiro, P. R. 1986;Pub. A. S. P.;98;1014
 
\rf Shapiro, P. R. \& Giroux, M. 1987;ApJ (Letters);321;L107
 
\rf Sherman, R. D. 1980;ApJ;237;355
  
\rf Silk, J. I. 1977;ApJ;211;638
 
\rf Smoot, G. F. \etal 1992;ApJ (Letters);396;L1-L18
 
\rf Stebbins, A. \& Silk, J. 1986;ApJ;300;1
 
\rf Steidel, C. C. \& Sargent, W. L. W. 1987;ApJ (Letters);318;L11
 
\rf Steidel, C. C. 1990;ApJ Suppl.;74;37
 
Teresawa, N. 1992, Preprint
 
\rf Tomisaka, K., Habe, H, \& Ikeuchi, S. 1980;Progr. Theor. 
Phys. (Japan);64;1587
 
\rf Weaver, R., McCray, R, Castor, J.,
Shapiro, P., \& Moore, R. 1977;ApJ;218;377
 
\rf Webb, J. K., Barcons, X., Carswell, R. F., \&
Parnell, H. C. 1992;MNRAS;255;319
 
\rf White, S. D. M., \& Rees, M. J. 1986;MNRAS;183;341


\clearpage

\clearpage

\clearpage

\clearpage

\clearpage

\clearpage

\begin{figure} 
\centerline{Figures~\ref{gpfig3a} and~\ref{gpfig3b}: IGM evolution.} 
Three different properties of the IGM 
(filling factor, temperature and ionization) are plotted as a function
of redshift for different choices of $M_c$, $f_g(5)$ and $\fcoll$.
$\Omega=1$, $\Omega_b=0.06$, $h=0.5$ and $f_m=0.1$ in the two
previous figures,
\ref{gpfig3a} and \ref{gpfig3b}.
In all panels, the different families of curves correspond to different
values of $f_g(5)$; 50\%, 20\%, 10\% and 1\% from left to right, with
the rightmost cases being omitted where they fail dismally.
In the porosity and temperature plots (the upper
two panels of 
\ref{gpfig3a} and \ref{gpfig3b}),
dashed lines correspond to $\fcoll=1$ and
solid ones to $\fcoll=0$, whereas only the
pessimistic $\fcoll=0$ case is plotted in the ionization plots (the lower
third). 
In the temperature and ionization plots 
(the lower two panels), the three branches of each curve correspond
to the three reheating scenarios: $f_{inj} = 0$,
$f_{inj} = 0.001$ and $f_{inj} = 0.02$.
\efig

\end{document}